\documentclass[twocolumn,aps,prl,bibnotes10pt,superscriptaddress]{revtex4-1}
\usepackage{inputenc}
\usepackage{mathtools}
\usepackage{amsmath}
\usepackage{amssymb}
\usepackage{graphicx}
\usepackage[dvipsnames]{xcolor}
\usepackage{ulem}
\usepackage {hyperref}
\hypersetup{
colorlinks=true,
citecolor=blue,
linkcolor=red,
urlcolor=blue
 , bookmarks=true,
 pdfmenubar=true
}
 
 \newcommand{\stkout}[1]{\ifmmode\text{\sout{\ensuremath{#1}}}\else\sout{#1}\fi}

\makeatletter

\DeclareTextSymbolDefault{\textquotedbl}{T1}

\usepackage{bm}
\usepackage{psfrag}
\usepackage{latexsym}
\usepackage{exscale}
\usepackage{float}

\makeatother

\begin{document}
\global\long\def\sgn{\mathrm{sgn}}%
\global\long\def\ket#1{|#1\rangle}%
\global\long\def\bra#1{\langle#1|}%
\global\long\def\sp#1#2{\langle#1|#2\rangle}%
\global\long\def\abs#1{\left|#1\right|}%

\title{Quantifying Computational Advantage of Grover's Algorithm with the Trace Speed}

\author{Valentin~Gebhart}\thanks{\href{mailto:gebhart@lens.unifi.it}{gebhart@lens.unifi.it} }

\affiliation{QSTAR, INO-CNR and LENS, Largo Enrico Fermi 2, I-50125 Firenze, Italy}
\affiliation{Universit\`a degli Studi di Napoli Federico II, Via Cinthia 21, I-80126 Napoli, Italy}

\author{Luca~Pezz\`e}\thanks{\href{mailto:luca.pezze@ino.it}{luca.pezze@ino.it} }
\affiliation{QSTAR, INO-CNR and LENS, Largo Enrico Fermi 2, I-50125 Firenze, Italy}

\author{Augusto~Smerzi}\thanks{\href{mailto:augusto.smerzi@ino.it}{augusto.smerzi@ino.it}}
\affiliation{QSTAR, INO-CNR and LENS, Largo Enrico Fermi 2, I-50125 Firenze, Italy}

\date{\today}

\begin{abstract}
Despite intensive research, the physical origin of the speed-up offered by quantum algorithms remains mysterious. 
No general physical quantity, like, for instance, entanglement, can be singled out as the essential useful resource. 
Here we report a close connection between the trace speed and the quantum speed-up in Grover's search algorithm implemented with pure and pseudo-pure states. 
For a noiseless algorithm, we find a one-to-one correspondence between the quantum speed-up and the polarization of the pseudo-pure state, which can be connected to a wide class of quantum statistical speeds.
For time-dependent partial depolarization and for interrupted Grover searches, the speed-up is specifically bounded by the maximal trace speed that occurs during the algorithm operations. 
Our results quantify the quantum speed-up with a physical resource that is experimentally measurable and related to multipartite entanglement and quantum coherence.
\end{abstract}

\maketitle

    Understanding and quantifying the key resource for the speed-up of quantum computations \cite{nielsen2002quantum,kaye2007introduction} has been a highly disputed topic over the past few decades \cite{vedral2010elusive}. 
    There has been particular interest in the role played by entanglement \cite{jozsa1997entanglement,ekert1998,Lloyd1999,linden2001good,jozsa2003role,Vidal2003,Biham2004,kenigsberg2006quantum,Horodecki2009,VandenNest2013}. 
    It is known that exponential speed-up of quantum algorithms implemented with pure states require multipartite entanglement \cite{jozsa2003role,Vidal2003}. 
    However, it was shown that a polynomial advantage can be achieved without entanglement \cite{bernstein1997quantum}. 
    Also, it is an open question whether exponential quantum advantage can be reached in mixed-state algorithms in absence of entanglement. 
    In this case, other quantum correlations such as quantum discord have been indicated as possible candidates for computational resources \cite{Datta2008,vedral2010elusive}. 
    Furthermore, it was shown that several entanglement measures cannot quantify advantages of many quantum algorithms \cite{VandenNest2013}. 
    Other possible resources have been considered such as coherence \cite{Hillery2016,Ma2016,Matera2016}, distinguishability \cite{vedral2010elusive}, contextuality \cite{Howard2014}, tree size \cite{cai2015} and interference \cite{Stahlke2014}.
    In short, despite having been the subject of extensive research, understanding the resource of the speed-up in quantum computations, even for a benchmark algorithm such as Grover's algorithm, is still an open and urgent quest.

	Quantum statistical speeds \cite{Wootters1981,Petz1996,Spehner2014,Braunstein1994,Gessner2018} offer a possible approach to quantify useful resources in quantum technology tasks. 
	As a major example, the quantum Fisher information \cite{helstrom1976quantum,Braunstein1994}, which is the quantum statistical speed associated with the Bures distance \cite{Braunstein1994}, was shown to fully characterize metrologically useful entanglement \cite{pezze2009,hyllus2012fisher,toth2012}, that is, the entanglement necessary for sub-shot-noise phase estimation sensitivities \cite{Pezze2016,Toth2014}. One might conjecture that different statistical speeds may be useful to characterize the performances of different quantum tasks. 
	Here, we use the trace speed ($\operatorname{TS}$), namely, the statistical speed associated to the trace distance \cite{nielsen2002quantum,Petz1996}, to quantify the speed-up in Grover's algorithm \cite{grover1997quantum} in both absence and presence of dephasing. 
	In particular, we show that in the pseudo-pure model without dephasing, the speed-up is completely determined by the polarization of the pseudo-pure state, which can be linked to a wide class of quantum statistical speeds.
	For general pseudo-pure dephasing models \cite{braunstein2000speed}, we prove that the maximal $\operatorname{TS}$ occuring during the algorithm bounds the speed-up, rendering it a necessary resource for quantum advantage. 
	The $\operatorname{TS}$ is an experimentally relevant measure of quantum coherence (asymmetry) \cite{marvian2014extending,Streltsov2017} and witnesses multipartite entanglement \cite{Gessner2018}. 
	To our knowledge, this is the first result for a physical resource in Grover's algorithm that generalizes to mixed state versions. This can pave the way to a new approach to investigate useful resources in quantum computations.

\section*{Results}

\subsection*{Grover's algorithm and its cost}

    Grover's search algorithm \cite{grover1997quantum} is one of the most important protocols of quantum computation \cite{nielsen2002quantum,kaye2007introduction}. 
    It searches an unstructured database of $N$ elements for a target $\omega$. 
    The target is marked in the sense that one is given a test function $f$ that vanishes for all elements but $\omega$. 
    The task is to identify $\omega$ with as few function calls as possible. 
    In the quantum version of the algorithm, a function call can be used as a measurement or as an application of a corresponding unitary, the so-called oracle unitary. 
    As we will discuss shortly, Grover's algorithm admits a quadratic advantage to classical search algorithms. 
    To utilize the exponential size of the dimensionality of composite quantum systems \cite{Lloyd1999}, we encode all different elements $x$ of the register into computational basis states of $n=\log_2 N$ qubits, $x\in \left\{0,1\right\}^{ n}$. 
    Grover's algorithm is performed by preparing the system in the register state $\left| \psi_\mathrm{in} \right\rangle=1/\sqrt{2^n}\sum_x \left| x \right\rangle$, where $\left| x \right\rangle$ are the computational basis vectors, followed by $k$ applications of the Grover unitary $G=U_d U_\omega$. 
    Here, the oracle unitary $U_\omega=1-2\left| \omega \right\rangle\left\langle \omega \right|$ represents a function call and the Grover diffusion operator is defined as $U_d=2\left| \psi_\mathrm{in}  \right\rangle\left\langle \psi_\mathrm{in}  \right|-1$. 
    After $k$ iterations of the Grover unitary, the state of the system is given by \cite{kaye2007introduction}
    \begin{equation}
    \left| \psi_k \right\rangle = \sin[(2k+1)\theta]\left| \omega \right\rangle+\cos[(2k+1)\theta]\left| \omega^\perp \right\rangle,\label{eq:pureevolution}
    \end{equation}
    where $\theta=\arcsin(1/\sqrt{2^n})$ and $\left| \omega^\perp \right\rangle=1/\sqrt{2^n-1}\sum_{x\neq\omega}\left| x \right\rangle$ is the projection of the initial state on the  subspace orthogonal to $\left| \omega \right\rangle$. 
    This yields a probability $p_k$ of finding the target state as $p_k=\sin^2 [(2k+1)\theta]$. 
    After $k_\mathrm{Gr}\approx (\pi/4) \sqrt{2^n}$ iterations one finds the target state $\left| \omega \right\rangle$ with probability $p_{k_\mathrm{Gr}}=1-\mathcal{O}(1/2^n)$ \cite{kaye2007introduction}.

	One defines the cost $C$ for a general search algorithm as the average number of applications of the test function $f$ (or its corresponding oracle unitary) required to find the target state \cite{braunstein2000speed}. Simply counting the oracle applications is also known as query complexity \cite{kaye2007introduction}, while other complexities such as gate complexity are usually not considered in Grover's algorithm (see \cite{Lloyd1999} for a discussion). 
	
	In the classical search algorithm, the query application can be thought of as opening one of $2^n$ boxes, where each box represents one state of the register. 
	For an unstructured search algorithm, i.e., in each iteration one randomly opens one of the $2^n$ boxes, the average number of steps needed to find the target state is given by $C_\mathrm{cl}=2^n$. 
	If one remembers the outcome of all previous searches, the cost can be reduced to $C_\mathrm{cl}=2^{n}/2+\mathcal{O}(1)$ \cite{braunstein2000speed}.	
	Note that $C_\mathrm{cl}$ for both structured and unstructured searches scales with $2^n$.

	In a quantum search algorithm, one uses $k$ oracle unitaries and a final oracle measurement yielding the target with probability $p_k$, such that the cost is given by \cite{braunstein2000speed}
	\begin{equation}
	C_\mathrm{qu}(k)=\dfrac{k+1}{p_k}. \label{eq:cost}
	\end{equation}
	Hence, the optimal cost is obtained by minimizing $C_\mathrm{qu}(k)$ over the number of oracle applications, $C_\mathrm{qu}=\min_k(k+1)/(p_k)$. 
	Let us emphasize that this definition of the cost does not distinguish between applying the oracle as a unitary or as a measurement observable. 
	
    In Grover's algorithm, the cost function Eq. (\ref{eq:cost}) is not necessarily minimal for the highest success probability $p_k$ of one single search \cite{zalka1999}. 
    However, the optimal number of steps $\tilde{k}_\mathrm{Gr}$ and the optimal cost $C_\mathrm{qu}$ for large $n$ still scales as $\tilde{k}_\mathrm{Gr}= r \sqrt{2^n}$ and $C_\mathrm{qu}=K \sqrt{2^n}$, where $r$ is the solution of $\tan(2r)=4r$ and $K=r/\sin^2(2r)$, yielding the quadratic speed-up over $C_\mathrm{cl}$. It was shown that this speed-up is optimal \cite{bennett1997strengths,zalka1999}. 
	 
    Grover's algorithm can be executed on a single multimode system and, therefore, simply makes use of superposition and constructive interference \cite{Lloyd1999,Caves2004,bhattacharya2002implementation}. 
    However, in order to reduce exponential overhead in space, time or energy, one usually considers a system composed of many qubits \cite{Lloyd1999,Caves2004}. 
	In this case, different measures of bipartite and multipartite entanglement have been used to detect entanglement during Grover's algorithm \cite{bruss2011multipartite,Meyer2002,Fang2005,Rungta2009,Rossi2013}. 
	Genuine multipartite entanglement was shown to be present already after the first step of the noiseless algorithm \cite{bruss2011multipartite}. 
	However, the quantitative relationship between these measures and speed-up was not resolved. 
	In particular, the methods could not be easily applied to any mixed state generalization of Grover's algorithm. 
	Quantum coherence \cite{Shi2017,anand2016coherence,Pan2019} and quantum discord \cite{Shi2017} have been considered as resources in the noiseless algorithm as well. 

\subsection*{Quantifying speed-up in the unitary algorithm}

    In this section, we quantify the speed-up in a mixed-state generalization of Grover's algorithm with quantum statistical speeds. 
    We consider Grover's algorithm with the register initialized in a pseudo-pure state \cite{havel1997ensemble}, while the algorithm is still implemented with unitary operations. 
    For a pure $n$-qubit state $\left| \psi \right\rangle$, the corresponding pseudo-pure state $\rho_{\psi,\epsilon}$ with polarization $\epsilon$ is defined as 
    \begin{equation}
    \rho_{\psi,\epsilon}=\epsilon\left| \psi \right\rangle \left\langle \psi \right| +\dfrac{1-\epsilon}{2^n}\mathbb{I}.\label{eq:pseudopure}
    \end{equation}
	Pseudo-pure states represent one of the simplest models for mixed-state quantum computation and play a central role in the partial depolarizing noise model we consider later.
	We replace the pure initial state $\left| \psi_\mathrm{in} \right\rangle$ with the pseudo-pure state $\rho_{\psi_\mathrm{in},\epsilon}$ such that, after $k$ Grover iterations, the state of the system is given by $\rho_k=\epsilon\left| \psi_k \right\rangle \left\langle \psi_k \right|+(1-\epsilon)\mathbb{I}/2^n$, with $\left| \psi_k \right\rangle$ defined in Eq. (\ref{eq:pureevolution}). The probability $p_k$ of finding the target state after $k$ steps is $p_k=\epsilon\sin^2 ((2k+1)\theta)+(1-\epsilon)/2^n$. 
	Here we observe that for $\epsilon=\mathcal{O}(1/2^n)$, it becomes more efficient to just measure the state without any iteration because the probability contribution due to the Grover iteration is no longer dominant \cite{linden2001good}. 
	However, if $\epsilon$ does not decrease exponentially with $n$, one can neglect the second term in $p_k$. 
	Hence, the minimum of Eq. (\ref{eq:cost}) occurs after the same number of steps as in the pure state algorithm while its minimal value $C_\mathrm{qu}$ is simply $C_\mathrm{qu}=C_\mathrm{qu,pure}/\epsilon$, where $C_\mathrm{qu,pure}$ is the cost of the pure state algorithm. 
	
	In this model, the speed-up of the algorithm is completely determined by the depolarization $\epsilon$. Therefore, all quantities which can be connected to $\epsilon$ are quantifiers of the speed-up. In particular, this holds for a general quantum statistical speed $\operatorname{QS}$ with the property that for $\epsilon \gg 1/2^n$, $\operatorname{QS}(\rho_{\psi,\epsilon})=\epsilon \operatorname{QS}(\left| \psi \right\rangle \left\langle \psi \right|)$. 
	Note that this property holds for a wide class of quantum statistical speeds, for instance, the generalized quantum Fisher information and the quantum Schatten speeds \cite{Gessner2018}. 	
	Given the relations $\epsilon= \operatorname{QS}(\rho_{\psi,\epsilon})/ \operatorname{QS}(\left| \psi \right\rangle \left\langle \psi \right|)$ and $C_\mathrm{qu}=C_\mathrm{qu,pure}/\epsilon=K\sqrt{2^n}/\epsilon$, we directly obtain the dependence of the cost function on the maximal $\operatorname{QS}$ as 
    \begin{equation}
	C_\mathrm{qu}(n,\operatorname{QS}_\mathrm{max})=K\sqrt{2^n}\frac{\operatorname{QS}^\mathrm{pure}_\mathrm{max}}{\operatorname{QS}_\mathrm{max}}, \label{eq:costpseudo}
    \end{equation}
	where $K=r/\sin^2(2r)\approx 0.69$ with $r$ being the solution of $\tan(2r)=4r$, and $\operatorname{QS}^\mathrm{pure}_\mathrm{max}$ being the maximal $\operatorname{QS}$ during the pure-state algorithm. 
	The quantum speed-up $S=C_\mathrm{cl}/C_\mathrm{qu}$ is thus given in terms of $\operatorname{QS}_\mathrm{max}$ as 
	\begin{equation}
    S= \frac{\sqrt{2^n}}{2 K}\frac{\operatorname{QS}_\mathrm{max}}{\operatorname{QS}^\mathrm{pure}_\mathrm{max}}\label{eq:speeduppseudo}.
    \end{equation}
    As we discuss in more detail in the Methods section, for $\epsilon$ below a critical polarization $\epsilon_c=K/\sqrt{2^n}$, a classical search becomes advantageous. For $\epsilon>\epsilon_c$, the above results are valid. Also, for small values of $n$, a rigorous computation of the cost has to be performed.
    
    As an example of a quantum statistical speed that will play an important role in the next section, consider the specific case of the trace speed ($\operatorname{TS}$). 
    The $\operatorname{TS}$ is the susceptibility of a quantum state $\rho$ to unitary displacements generated by a generic Hamiltonian $H$ \cite{marvian2014extending}. 
    That is, the $\operatorname{TS}$ quantifies the distinguishability between $\rho$ and $\rho(t)=e^{-iHt}\rho e^{iHt}$ for small $t$. It is defined as \cite{nielsen2002quantum,marvian2014extending,Gessner2018}
	\begin{equation}
    \operatorname{TS}(\rho,H) = \left\lVert \partial_t \rho(t) \big|_{t=0} \right\rVert_1  = \left\lVert \left[\rho, H \right] \right\rVert_1 ,
    \end{equation}
    where $[\cdot,\cdot]$ is the commutator and $\left\lVert \cdot \right\rVert_1$ is the $l_1$-norm, defined as $\left\lVert A \right\rVert_1 = \operatorname{tr}\left[ \sqrt{A^\dag A} \right]$ for a generic operator $A$. Since $\operatorname{TS}(\rho_{\psi,\epsilon})=\left\lVert \left[\rho_{\psi,\epsilon}, H \right] \right\rVert_1=\left\lVert \epsilon\left[\left| \psi \right\rangle \left\langle \psi \right|, H \right]+(1-\epsilon)/2^n\left[ \mathbb{I} , H \right] \right\rVert_1=\epsilon \operatorname{TS}(\left| \psi \right\rangle \left\langle \psi \right|)$, the $\operatorname{TS}$ can be used as the $\operatorname{QS}$ in Eq. (\ref{eq:speeduppseudo}). 
	
	In general, $\operatorname{TS}$ is a measure of coherence, in this case usually referred to as asymmetry \cite{marvian2014extending}: a state with no coherence with respect to $H$, namely a classical mixture of its eigenstates, will not change under phase displacements, while off-diagonal matrix elements (coherences) of $\rho$ are responsible for a finite susceptibility to phase displacements. 
	The $\operatorname{TS}$ is upper bounded by the quantum Fisher information \cite{Petz1996}. 
	If the system is a composite system of $n$ qubits and $H$ is the sum of local Hamiltonians $H_i$, $H=\sum_{i=1}^n H_i$ with $\operatorname{spec}(H_i)=\left\{-1/2,1/2\right\}$ and $\operatorname{TS}(\rho,H)>\sqrt{nr}$, it follows that $\rho$ has to be at least $(r+1)$-partite entangled \cite{hyllus2012fisher,toth2012,Gessner2018}, namely, it cannot be written as a mixture of $r$-producible pure states. A pure state is $r$-producible if it is a tensor product of subsystems with each subsystem containing at most $r$ qubits. 
	Since the value of $\operatorname{TS}$ depends on the generating Hamiltonian $H$, we consider the optimization over all Hamiltonians of the above form. When the whole evolution is restricted to the completely symmetric subspace, it suffices to perform this optimization over collective spin Hamiltonians, $H_i=\mathbf{n}\cdot\boldsymbol{\sigma}^{(i)}/2$, where $\mathbf{n}$ is a point on the unit sphere and $\boldsymbol{\sigma}^{(i)}$ are the Pauli operators for the $i$-th qubit. 
	For pure states $\left| \psi \right\rangle$, the optimized $\operatorname{TS}$ coincides with the square root of the largest eigenvalue of the matrix
	$\Gamma_{ij}=4\left(\operatorname{Re}[\left\langle J_iJ_j\right\rangle]-\left\langle J_i\right\rangle\left\langle J_j\right\rangle\right)$ \cite{hyllus2012fisher}. 
	Here, $J_m=\sum_{i=1}^n\mathbf{e}_m\cdot\sigma^{(i)}/2$ is the coherent spin operator in $\mathbf{e}_m$-direction, $m=x,y,z$, and $\langle \cdot \rangle$ is the expectation value with respect to the state $\left| \psi \right\rangle$.
	
\begin{figure}
		\centering
		\includegraphics[width=\linewidth]{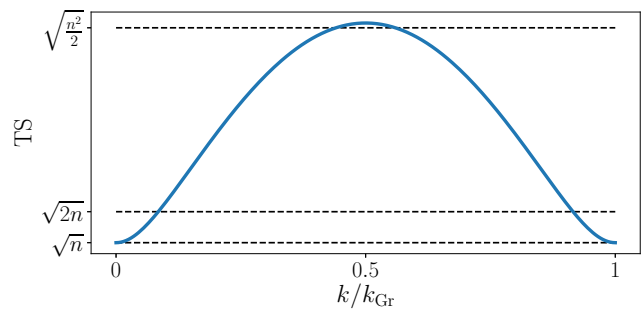}
		\caption{The dependence of the optimized trace speed $\operatorname{TS}$ on the iteration step $k$ in the pure state Grover's algorithm (solid line). The dashed lines indicate thresholds above which $\operatorname{TS}$ detects bipartite ($\sqrt{n}$), three-partite ($\sqrt{2n}$) and ($n/2+1$)-partite ($\sqrt{n^2/2}$) entanglement. Here, $n=30$, $k_\mathrm{Gr}\approx(\pi/4)\sqrt{2^n}$.}
		\label{fig:QFIpure}
	\end{figure}

	Let us first discuss the $\operatorname{TS}$ for the standard version of Grover's algorithm implemented with pure states and unitary evolution, as introduced above. 
	Without loss of generality, we consider $\left| \omega \right\rangle=\left| 0 \right\rangle^{\otimes n}$. This choice corresponds merely to a relabeling of the computational basis vectors of each qubit. Importantly, by this relabeling, the dynamics of the Grover search, as well as the optimized $\protect \operatorname {TS}$ of the state during the evolution, are not altered, while the calculation of $\protect \operatorname {TS}$ is highly facilitated. 
	Since $\left| \psi_\mathrm{in} \right\rangle$ and $\left| 0 \right\rangle^{\otimes n}$ are elements of the completely symmetric subspace and $G$ commutes with all permutations of the qubits, the complete evolution is restricted to the symmetric subspace, facilitating the computation of $\operatorname{TS}$. 
	By neglecting terms in $\mathcal{O}(1/2^n)$, one can exactly compute the largest eigenvalue of $\Gamma_{ij}$ at any step $k$, yielding the optimized $\operatorname{TS}$, see Methods for details. 
	In Fig. (\ref{fig:QFIpure}), we show the optimized $\operatorname{TS}(k)$ for $n=30$ qubits. 
	The initially separable state $\left| \psi_\mathrm{in} \right\rangle$ evolves into a multipartite entangled state already after the first oracle operation. 
	Multipartite entanglement further increases until reaching a maximal value of 
	\begin{equation}
	\operatorname{TS}^\mathrm{pure}_\mathrm{max}=\sqrt{\dfrac{n(n+1)}{2} }
	\end{equation}
	which occurs at $k=k_\mathrm{Gr}/2$. This detects $(n/2+1)$-partite entanglement during the pure state Grover algorithm. For $k>k_\mathrm{Gr}/2$, multipartite entanglement detected by the $\operatorname{TS}$ decreases until the algorithm reaches the separable target state $\left| \omega \right\rangle$.
	
	Since $\operatorname{TS}(\rho_{\psi,\epsilon},H)=\epsilon \operatorname{TS} (\left| \psi \right\rangle\left\langle \psi \right|,H)$, the maximal $\operatorname{TS}$ during the Grover algorithm using a pseudo-pure initial state with polarization $\epsilon$ is $\operatorname{TS}_\mathrm{max}=\epsilon \sqrt{n(n+1)/2}$. Hence, $\operatorname{TS}$ witnesses $\epsilon (n+1)/2$-partite entanglement. 
	Note that for $\epsilon<2/(n+1)$, $\operatorname{TS}$ does not detect entanglement anymore. 
	It was already observed that for polarizations $\epsilon>1/2^{n/2}$ the algorithm still offers a speed-up \cite{Biham2002,vedral2010elusive,Kay2015}, indicating that entanglement detected by $\operatorname{TS}$ is not necessary for quantum speed-up.

\subsection*{Trace speed and the algorithm under partial depolarization}

\begin{figure}
        \centering
        \includegraphics[width=\linewidth]{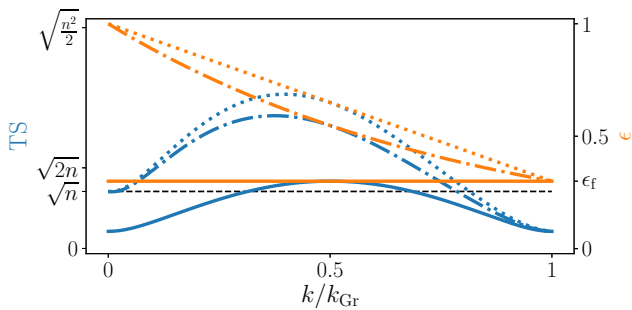}
        \caption{Trace speed during the pseudo-pure version of Grover's algorithm. Polarizations $\epsilon(k)$ (orange lines) and trace speeds $\operatorname{TS}(k)$ (blue lines) for an initial pseudo-pure state without dephasing (solid), an initial pure state with linearly decaying polarization (dotted) and an initial pure state with exponentially decaying polarization (dash-dotted). Here, $n=30$, $\epsilon_\mathrm{f}=0.3$, $k_\mathrm{Gr}\approx(\pi/4)\sqrt{2^n}$.}
        \label{fig:QFIdeph}
	\end{figure}

    The results of pseudo-pure initial states can be generalized to search dynamics subject to time-dependent partial depolarization (see Refs. \cite{cohn2016grover,vrana2014fault} for earlier investigations). 
    In this case, the state after $k$ steps of the algorithm is given by 
	\begin{equation}
	\rho_k=\epsilon(k)\left| \psi_k \right\rangle \left\langle \psi_k \right|+\dfrac{1-\epsilon(k)}{2^n}\mathbb{I},\label{eq:pseudopureevolution}
	\end{equation}
	where the now time-dependent decreasing polarization $\epsilon(k)$ represents both initial impurity and partial depolarization during the algorithm. 
	The depolarization channel is a widely used noise model whenever the exact form of the noise is not known \cite{wilde2013}. As a worst case noise scenario, the knowledge of the state is completely erased with some probability.
	As can be seen in Fig. (\ref{fig:QFIdeph}), different polarization functions $\epsilon(k)$ with the same final polarization $\epsilon_\mathrm{f}$ can lead to different maximal $\operatorname{QS}$ during the iteration. 
	While the one-to-one correspondence between the $\operatorname{QS}$ and the speed-up is generally lost, as shown below, we can still bound the speed-up using the $\operatorname{TS}$.

	For a partial depolarization during the algorithm it turns out that, in general, it is optimal to stop the iterations and perform the final measurement already at earlier steps $k_\mathrm{int}<\tilde{k}_\mathrm{Gr}$ \cite{cohn2016grover}. We divide the examination into the cases $k_\mathrm{int}\leq k_\mathrm{Gr}/2$ and $k_\mathrm{int}\geq k_\mathrm{Gr}/2$, that is, whether we interrupt the iteration before or after the pure state algorithm would have already reached its maximal $\operatorname{TS}$, see Fig. (\ref{fig:QFIpure}). In the case $k_\mathrm{int}\geq k_\mathrm{Gr}/2$, the cost can be bounded by $C_\mathrm{qu}\geq K \sqrt{2^n}/\epsilon(k_\mathrm{int})$. This is because if one could completely stop the dephasing from this point, one could reduce the cost until reaching the optimal value of $K \sqrt{2^n}/\epsilon(k_\mathrm{int})$, see Eq. (\ref{eq:costpseudo}). Since $k_\mathrm{int}\geq k_\mathrm{Gr}/2$, we have $\epsilon(k_\mathrm{int})\leq \epsilon(k_\mathrm{Gr}/2)$ and $\operatorname{TS}(k_\mathrm{Gr}/2)\leq \operatorname{TS}_\mathrm{max}$ ($\operatorname{TS}_\mathrm{max}$ is the maximal $\operatorname{TS}$ before the interruption). Finally, using $\epsilon(k_\mathrm{Gr}/2)=\operatorname{TS}(k_\mathrm{Gr}/2)/\operatorname{TS}^\mathrm{pure}_\mathrm{max}$, one can then bound $C_\mathrm{qu}\geq K \sqrt{2^n}/\epsilon(k_\mathrm{Gr}/2) \geq (K \sqrt{2^n} \operatorname{TS}_\mathrm{max}^\mathrm{pure})/(\operatorname{TS}_\mathrm{max})$, yielding the following bound 
    \begin{equation}
	S \leq \frac{\sqrt{2^n}}{2 K}\frac{\operatorname{TS}_\mathrm{max}}{\operatorname{TS}^\mathrm{pure}_\mathrm{max}}.\label{eq:finalbound}
	\end{equation}
	
	The case $k_\mathrm{int}\leq k_\mathrm{Gr}/2$ corresponding to strong dephasing becomes more technical since, in the early regime, the maximal $\operatorname{TS}$ is not simply bounded by $\epsilon(k)\operatorname{TS}^\mathrm{pure}_\mathrm{max}$. 
	However, as we show in Methods, by using the explicit form of the $\operatorname{TS}$, the bound Eq. (\ref{eq:finalbound}) still holds. 
	At this point, the $\operatorname{TS}$ stands out from other quantum statistical speeds.
	For instance, the bound does not hold when using the quantum Fisher information as $\operatorname{QS}$. 
	These results for the case of an interruption of the iteration due to minimization of the cost can also be applied to the case of a general interruption of the iteration. 
	Stopping the algorithm at any time will yield an average speed-up which is always bounded by the maximal $\operatorname{TS}$ occurring before the interruption.

%%%%%%%%%%%%%%%%%%%%%%%%%%%%%%%%%%%%%%%%%%%%%%%%%%%%%%%%%%%%%%%%%%%%%%%%%%%%%%%%%%%%%%%%%%%%%%%%%%%%%%%%%%%%%%%%%%%%%%%%%%%%%%%%%%%%%%%%%%%%%%%%%%%%%%%%%%%%%%%%%%%%%%%%%%%%%%%%%%%%%%%%%
%%%%%%%%%%%%%%%%%%%%%%%%%%%%%%%%%%%%%%%%%%%%%%%%%%%%%%%%%%%%%%%%%%%%%%%%%%%%%%%%%%%%%%%%%%%%%%%%%%%%%%%%%%%%%%%%%%%%%%%%%%%%%%%%%%%%%%%%%%%%%%%%%%%%%%%%%%%%%%%%%%%%%%%%%%%%%%%%%%%%%%%%%

\section*{Discussion and conclusions}

	To summarize, we showed that both in the pure state version of the Grover search algorithm and a general pseudo-pure generalization, the trace speed ($\operatorname{TS}$) can be used to quantify and bound the possible speed-up over a classical search. 
	These results offer an unprecedented connection between the speed-up in Grover's algorithm and a physical resource beyond the case of ideal, noiseless quantum algorithms. 
	The $\operatorname{TS}$ relates the computational speed of Grover's algorithm to both multipartite entanglement and quantum coherence. 
    It should be noticed that the relation with multipartite entanglement depends on the $n$-qubit implementation that we have considered, while the algorithm can also be implemented with a single $2^n$-level system \cite{Lloyd1999}. 
    Indeed, as mentioned above, the operating principle of the algorithm and the number of queries used (which determines the cost) do not depend on which implementation we use. 
    Therefore, multipartite entanglement cannot be considered as the key resource for the quantum speed-up.
    We thus argue that the correct interpretation of our result is the evidence that the resource for speed up in query complexity is quantum coherence as captured by the $\operatorname{TS}$. 
    However, multipartite entanglement is crucial to reduce other costs such as space or energy \cite{Lloyd1999}. 
    We point out that the interpretation of the $\operatorname{TS}$ as quantum coherence holds for any implementation of the algorithm.

    The role of quantum coherence during the noiseless Grover's algorithm has already been investigated in Refs. \cite{Shi2017,anand2016coherence}. 
    These works found a one-to-one correspondence between the $l_1$-norm of coherence which is decreasing during the algorithm and the increasing success probability. 
    Both approaches have not been generalized to mixed state versions of the algorithm. 
    In our case, a different measure of coherence, namely the $\operatorname{TS}$, is connected to the average cost of the algorithm. 
    It reaches its maximal value during the algorithm and offers a physical resource also for pseudo-pure generalizations. 
    In Refs. \cite{Shi2017,anand2016coherence},  the $l_1$-norm of coherence and  the relative entropy of coherence are used which detect different states as highly coherent as $\operatorname{TS}$ would. 
    For instance, while the $l_1$-norm detects the initial state $\left| \psi_\mathrm{in} \right\rangle=1/\sqrt{2^n}\sum_x \left| x \right\rangle$ as maximally coherent, $\operatorname{TS}$ would detect $(\left| 0 \right\rangle^{\otimes n}+\left| 1 \right\rangle^{\otimes n})/\sqrt{2}$ as maximally coherent. 
    For a discussion of these so-called speakable and unspeakable coherence, see for instance Ref. \cite{Marvian2016}. 

    Finally, we emphasize that the $\operatorname{TS}$ can be measured or efficiently bounded experimentally. 
    Following Refs. \cite{Strobel2014,Pezze2016a}, one measures the Kolmogorov distance between the probability distribution of $\rho(0)$ and $\rho(t)$, for a given measurement observable. 
    A quadratic series expansion of the Kolmogorov distance for sufficiently small $t$ yields the Kolmogorov speed which is a lower bound to the TS and depends on the considered measurement observable. 
    The TS is obtained by maximizing the Kolomogorov speed over all possible observables \cite{nielsen2002quantum}.

    To conclude, the analysis of the $\operatorname{TS}$ might inspire further investigations of the still unanswered search for the origins and quantification of quantum advantage. 
    In particular, one could check the importance of the $\operatorname{TS}$ and other quantum statistical speeds for other oracle-based quantum algorithms such as, e.g., the Deutsch-Jozsa algorithm or Simon's algorithm, or general quantum technology tasks. 
    Also, whether or not the $\operatorname{TS}$ is a necessary resource in different noisy variations of Grover's algorithm, merits further investigation. 
    More general dephasing models or unitary noise could be considered that render the analysis more cumbersome.
    Overall, our results suggest that quantum statistical speeds can be used to recognize useful properties of quantum states for different quantum technology tasks.

\section*{Methods}

\subsection*{Cost dependence for small polarizations}
As discussed in the main text, for initial polarizations $\epsilon \sim 1/2^{n}$, a classical search is less costly than performing the Grover iterations. 
Here, we discuss the behavior of the cost in this regime. 
We observe numerically that for polarizations $\epsilon$ above a critical value $\epsilon_c$, the minimum of the cost Eq. (\ref{eq:cost}) is always obtained after $\tilde{k}_\mathrm{Gr}= r \sqrt{2^n}$ iteration steps, where the cost is given by $C_\mathrm{qu}=K\sqrt{2^n}/\epsilon$ ($K\approx 0.69$, see main text). For $\epsilon<\epsilon_c$, the cost is minimized for $k=0$ iterations, that is, performing a classical search with $C_\mathrm{qu}=2^n$. The exact value of $\epsilon_c$ is given by the equality of the costs, $K\sqrt{2^n}/\epsilon=2^n$, yielding 
\begin{equation}
\epsilon_c=\frac{K}{\sqrt{2^n}}.
\end{equation}
Note that for small $n$, the minimization of the cost has to be performed on a discrete grid. Therefore, a case study for each $n$ has to be performed. 

\subsection*{Optimized trace speed during the pure-state algorithm}
Here, we describe the derivation of the analytical formula for the optimized trace speed during the algorithm. 
Since the evolution of the algorithm is restricted to the symmetric subspace, the optimized trace speed $\operatorname{TS}$ during the algorithm is given by the square root of the largest eigenvalue of $\Gamma_{ij}=4\left(\operatorname{Re}[\left\langle J_iJ_j\right\rangle]-\left\langle J_i\right\rangle\left\langle J_j\right\rangle\right)$ \cite{hyllus2012fisher}, where the expectation value $\langle \cdot \rangle$ is computed with respect to the state state $\left| \psi_k \right\rangle = \sin[(2k+1)\theta]\left| \omega \right\rangle+\cos[(2k+1)\theta]\left| \omega^\perp \right\rangle$, cf. Eq. (\ref{eq:pureevolution}). The computation is straightforward, for instance, for $J_x =\sum_{i=1}^n\sigma^{(i)}_x/2$, one finds that
\begin{align}
\langle J_x \rangle &= \frac{1}{2}\sum_{i=1}^n\left\langle \psi_k\right| \sigma^{(i)}_x \left| \psi_k\right\rangle\notag \\ 
&= \frac{n}{2}\left\langle \psi_k\right| \sigma^{(1)}_x \left| \psi_k\right\rangle\notag \\
%& = \frac{n}{2} \cos^2[(2k+1)\theta] \left\langle \omega^\perp \right| \sigma^{(1)}_x\left| \omega^\perp \right\rangle+ \mathcal{O}(1/2^n) \notag\\ 
&= \frac{n}{2} \cos^2[(2k+1)\theta] + \mathcal{O}(1/2^n). 
\end{align}
Eventually, we obtain 
\begin{widetext}
\begin{equation}
\Gamma = \begin{pmatrix}
n+n(n-1)\cos^2\theta_k- n^2 \cos^4\theta_k & 0 & -n^2 \sin^2\theta_k\cos^2\theta_k\\
0 & n & 0 \\
-n^2 \sin^2\theta_k\cos^2\theta_k & 0 & n+n(n-1)\sin^2\theta_k - n^2 \sin^4\theta_k, 
\end{pmatrix}+ \mathcal{O}(1/2^n),
\end{equation}
where we defined $\theta_k = (2k+1)\theta$.
Taking the square root of the largest eigenvalue and neglecting terms of order $\mathcal{O}(1/2^n)$, we find the optimized $\operatorname{TS}$ after the $k$-th step of the pure-state algorithm as 
\begin{equation}
 \operatorname{TS}_\mathrm{pure}(k) =\sqrt{\frac{1}{8}n\left(4+n-f(k)n+\sqrt{8[1+f(k)]+n^2[1-f(k)]^2}\right)} + \mathcal{O}(1/2^n). \label{eq:TSanalytic}
\end{equation}
\end{widetext}
with $f(k)=\cos[4(2k+1)\theta]$.

\subsection*{Partial depolarization}

Here, we discuss the case of an interruption of the search algorithm at an early step $k\leq k_\mathrm{Gr}/2$, i.e., at a step where the pure state algorithm would not have reached its maximal $\operatorname{TS}$ yet. The cost for stopping the algorithm at step $k$ is given by $C_\mathrm{qu}(k)=(k+1)/(\epsilon(k)\sin^2((2k+1)\theta))$, see Eq. ($1$) in the main text. The $\operatorname{TS}$ at step $k$ still fulfills $\operatorname{TS}(k)=\epsilon(k)\operatorname{TS}_\mathrm{pure}(k)$, where $\operatorname{TS}_\mathrm{pure}(k)$ is the $\operatorname{TS}$ in the pure state algorithm, see Fig. ($1$) in the main text. Therefore, we have 
\begin{align}
C_\mathrm{qu}(k)&=\frac{k+1}{\sin^2 [(2k+1)\theta]}\frac{\operatorname{TS}_\mathrm{pure}(k)}{\operatorname{TS}(k)}\notag \\ 
&\geq \frac{k+1}{\sin^2 [(2k+1)\theta]}\frac{\operatorname{TS}_\mathrm{pure}(k)}{\operatorname{TS}_\mathrm{max}} \notag \\ 
&=\frac{a(k)}{\operatorname{TS}_\mathrm{max}}
\end{align}
where $\operatorname{TS}_\mathrm{max}$ is the maximal $\operatorname{TS}$ until the interruption step $k$ and we regrouped all other factors into $a(k)$. To further examine this expression, we use the exact form of the pure state $\operatorname{TS}$, $\operatorname{TS}_\mathrm{pure}(k)$, cf. Eq. (\ref{eq:TSanalytic}). We can then compare the factor $a(k)$ with the factor $b=K\sqrt{2^n}\sqrt{n(n+1)/2}$ from the current bound, Eq. ($9$) in the main text. By writing $x=k/\sqrt{2^n}$, one finds for large $n$ 
\begin{equation}
a(k)- b = \frac{n\sqrt{2^n}}{\sqrt{8}\sin^2(2x)}\left[K(\cos(4x)-1)+2x\sin4x\right].
\end{equation}
For $0\leq x\leq \pi/8$ ($0\leq k\leq \pi/8 \sqrt{2^n}  = k_\mathrm{Gr}/2$) and using $K\approx 0.69$, one finds that $a(k)-b>0$. Therefore, the bound of Eq. ($9$) still holds for the regime $k\leq k_\mathrm{Gr}/2$.

\subsection*{Acknowledgments}
The authors acknowledge financial support from the European Union’s Horizon 2020 research and innovation programme -- Qombs Project, FET Flagship on Quantum Technologies grant no. 820419. 

 \subsection*{Author Contributions}
V.G., L.P. and A.S. all contributed equally to this work.

  \subsection*{Competing interests}
The authors declare no competing interests.

\end{document}